\documentclass[11pt]{article}
\usepackage{graphicx}
\usepackage{amssymb}
\usepackage{amsmath}
\usepackage[square,numbers,sort&compress]{natbib}

\newcommand{\beq}{\begin{eqnarray}}
\newcommand{\eeq}{\end{eqnarray}}

\newcommand{\Slash}[1]{\ooalign{\hfil/\hfil\crcr$#1$}}

\def\gsim{\displaystyle\mathop{>}_{\sim}}

\begin{document}
\title{Magnetic interaction induced by the anomaly in kaon-photoproductions}
\author{S. Ozaki, H. Nagahiro and A. Hosaka\\
{\it Research Center for Nuclear Physics,  Osaka University,}\\
{\it Ibaraki 567-0047, Japan}}
\date{}
\maketitle
%
%======================
\begin{center}
{\bf Abstract}
\end{center}
We study the role of magnetic interaction in the photoproduction 
of the kaon and hyperon.  
We find that the inclusion of a higher order diagram 
induced by the Wess-Zumio-Witten term 
has a significant contribution to the magnetic amplitude, 
which is compatible to the observed photon asymmetry in the forward 
angle region.
This enables us to use the 
$K^*$ coupling constants which have been determined in a microscopic way
rather than the phenomenological ones which differ largely from the microscopic ones. 

%======================

\vspace*{1cm}

%=====================
%\section{Introduction}
%=====================

Strangeness production is one of important subjects in hadron 
and nuclear physics.  
It is the basis of hyperon interactions and hyper-nuclear 
physics, where an expansion to the new dimension of strange matter 
is being explored.  
Many reactions of producing exotic states including pentaquarks are
also associated with strangeness productions.  
Therefore, the understanding of the production mechanism 
is indeed a key to discuss the above interesting physics.  
However, the theoretical status of the production mechanism 
is not yet well established.  

Photoproduction of kaon and hyperon is one of the simplest 
reactions among 
them~\cite{Sumihama:2005er,Kohri:2006yx,Bradford:2005pt,Glander:2003jw}.  
If the kaon is treated as a light particle as the pion, which are 
altogether regarded as the Nambu-Goldstone bosons, 
their interactions are governed by the low energy 
theorems of chiral symmetry, respecting flavor 
SU(3) symmetry as well.  
In fact, in order to produce the kaon and hyperons,  
energy of order of 1 GeV must be deposited and therefore, 
the naive application of the low energy theorems might be 
doubtful.  

Yet, many reaction studies so far are based on the effective 
Lagrangian approach, the form of which is determined by 
symmetries compatible with flavor and chiral symmetries with much 
success~\cite{Lee:1999kd,Guidal:2003qs}.  
Then, various coupling constants such as 
kaon and vector $K^*$ coupling constants are treated 
as parameters.  
In literatures, the kaon coupling constants such as 
$g_{KN\Lambda}$ and $g_{KN\Sigma}$ are determined microscopically
from the pion coupling $g_{\pi NN}$ under flavor SU(3) symmetry rotations 
with suitable input of the $F/D$ ratio~\cite{Reuber:1993ip, Stoks:1999bz}.  
This method may be also applied 
to the $K^*$ coupling constants.  
If we have a consistent understanding for the strong interaction, such parameters should be 
universal and can be applied to other reactions. 

By now, many experimental data are available for the kaon photoproductions from various photon facilities 
including LEPS~\cite{Sumihama:2005er,Kohri:2006yx}, 
CLAS~\cite{Bradford:2005pt} and SAPHIA~\cite{Glander:2003jw}.
They provide detailed information on energy dependence of cross sections, angular dependence and  polarization phenomena.
In such a situation, it is very important to study these data  based on a microscopic description which
is compatible with QCD.

In the present paper, we would like to study the relevant reactions in the effective 
Lagrangian approach, where Born diagrams are computed as shown in Fig.1(a)-(d). There, various coupling constants
are input parameters, reflecting the microscopic nature of the interactions.
A unique feature of recent photoproduction experiments is in the use of the polarized photon.
In particular, the LEPS group has been providing data for the photon asymmetry in the forward angle region. Here we focus on the reaction
\beq
\gamma + p \to K^{+} + \Lambda
\eeq
in order to clarity the existing problem and to show its resolution.

\begin{figure}[tbp]
\centerline{
\includegraphics[width=130mm,bb=30 70 800 480,clip]{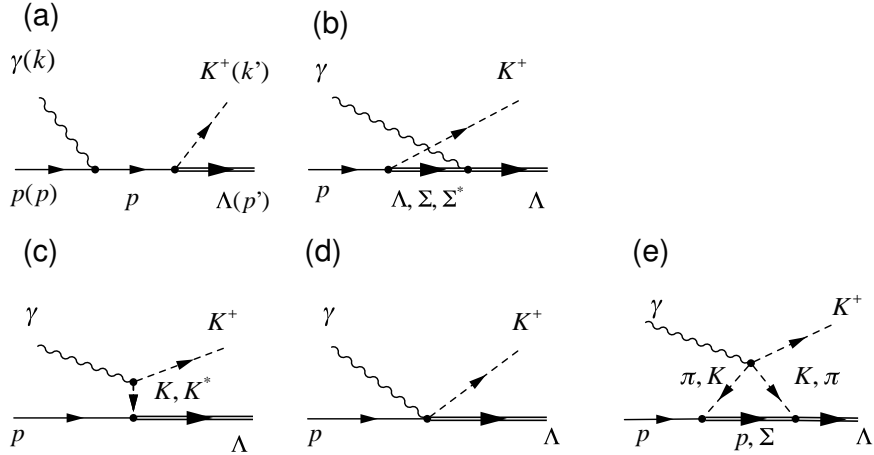}
}
\caption{\small Feynman diagrams for the kaon phtoproduciton.
Diagrams (a)-(d) are for the conventional Born diagrams for the $s, u, t$ and contact terms. The diagram (e) is
the one loop diagram induced by the WZW term.}
\label{diagrams}
\end{figure}

An advantage in the forward angle region is the $t$-channel dominance as shown in Fig.~\ref{diagrams}(c), where one can study 
the interactions of the exchanged particles, exclusively.
There we expect that dominant contributions are from $K$ and $K^*$ exchanges.   
The properties of these two meson exchanges associated with the electromagnetic interaction are of
interest and can be
distinguished by the azimuthal $\phi$-angler distribution by using the linearly polarized 
photon~\cite{Mibe:2005er}.
For productions of a pseudoscalar particle (in the present study it is the kaon), 
if the final state particles are produced more along the 
photon polarization direction, the interaction is 
dominated by the electric component and is induced by the 
$K$ exchange.  
In contrast, 
if the final state particles are produced more along the direction perpendicular to the 
polarization, the interaction is 
dominated by the magnetic component and is induced by the 
$K^*$ exchange~\cite{Nam:2006cx}.  

In order to characterize the $\phi$ angular distribution, 
the photon asymmetry is used, which is given by 
\beq
A = \frac{\sigma_\perp - \sigma_{//}}{\sigma_\perp + \sigma_{//}}  ,
\label{def_asymmetry}
\eeq
where $\sigma_\perp$ and $\sigma_{//}$ are defined in Refs~\cite{Mibe:2005er,Nam:2006cx}.
By this definition, the interaction is dominated by the electric component if $A$ is negative, while 
magnetic if positive. Observations including recent kaon photoproductions indicate
that the asymmetry $A$ is positive~\cite{Sumihama:2005er,Kohri:2006yx}.  
In order to explain the positive values,  a rather 
strong magnetic interaction is needed which has
been incorporated by the strong coupling of $K^*$ into the previous calculations.  
In Table~\ref{couplings} we show phenomenological coupling constants~\cite{Lee:1999kd} which are compared with those determined microscopically~\cite{Reuber:1993ip}. The former are determined to fit the data including the asymmetry, while
the latter are by SU(3) symmetry with the $F/D$ ratio guided by the vector meson dominance 
for $g^{V}_{K^*}$, and by SU(6) relation for $g^{T}_{K^*}$.  
We see that the phenomenological couplings (left) are typically five times larger than the microscopic
couplings (right). 

%-------------------------------------------------------
\begin{table}[tbp]
 \caption{Comparison of $K$ and $K^{*}$ coupling constants from Refs.~\cite{Lee:1999kd} and~\cite{Reuber:1993ip}. In Ref.~\cite{Lee:1999kd}, $K^{*} $ coupling constants are determined by phenomenologically in order to reproduce the photoproduction data, while in Ref.~\cite{Reuber:1993ip} they are constructed microscopically.} 
 \begin{center}
  \begin{tabular}{ c c c }
  \hline
 & Phenomenological & Microscopic \\ 
  \hline
$g_{KN\Lambda}$ & $-13.46$ & $-12.65$ \\
$g_{KN\Sigma}$ & $4.25$ & $5.92$ \\
$g^V_{K^*N\Lambda}$ & $-25.21$ & $-5.63$ \\
$g^T_{K^*N\Lambda}$ & $33.13$ & $-18.34$ \\
$g^V_{K^*N\Sigma}$ & $-15.33$ & $-3.25$ \\
$g^T_{K^*N\Sigma}$ & $-29.67$ & $7.86$ \\
 \hline
 \end{tabular}
 \end{center}
 \label{couplings}
\end{table}
%-------------------------------------------------------

In order to supply the missing strength of the magnetic interaction, 
in this paper, we would like to propose an additional mechanism 
induced by the Wess-Zumino-Witten (WZW) term associated with the 
QCD anomaly~\cite{Witten:1983tw}.  
In the presence of the gauged WZW term, there is a 
$\gamma MMM$ ($M$ = meson) vertex which may contribute 
to kaon photoproductions with one loop as shown in Fig.~\ref{diagrams}(e).  
There are several good features in the consideration of such a processe.  
First, the strength of the anomalous term of $\gamma MMM$ is 
unambiguously determined by QCD.  
It is given by 
\beq
\mathcal{L}_{\gamma K^{+} K^{-} \pi^{0}}
&=&\frac{2}{3}ieN_{c}\epsilon_{\mu\nu\sigma\rho}A^{\mu}\frac{1}{(2f_{\pi})^3\pi^2}\partial^{\nu}K^{+}\partial^{\sigma}K^{-}
\partial^{\rho}\pi^{0},
\eeq
where $e$ is the electric charge, $N_{c}=3$ the number of colors and $f_{\pi}=93\ \rm{GeV}$ the pion decay constant. Second, it contains the antisymmetric tensor $\epsilon_{\mu \nu \sigma \rho}$ ( $\epsilon^{01 2 3}=-\epsilon_{01 2 3}=+1$), and 
contributes to the magnetic interaction of the photon.  
Third, the anomalous vertex contains the incident 
photon momentum ($k$), and therefore, the contribution is 
expected to increase as the photon energy is increased.  
Of course, the amplitude should not keep increasing up to very large $k$, since it violates the unitarity. 
However, we expect that it should happen at low energy.  
Motivated by these considerations, we include the one-loop 
process as shown in Fig.~\ref{diagrams}(e) in the photoproduction in addition to the Born diagrams.  

There are two remarks in order.  
One is the fact that the loop integral diverges quadratically, 
and therefore we need to introduce a suitable regularization. 
Another is the problem of  double counting with the $K^*$ exchange, 
since $K^*$ could be regarded as a pair of $K$ and $\pi$.
In the present one loop diagram, however, there is no such $K^*$ component which are described as a correlated $K\pi$ pair.  Such a correlation is not included in the lowest one loop diagrams. 

Let us now turn to the formulation.  
The Born diagrams are calculated by the effective 
Lagrangian method. The interaction terms for the strong interactions are given by 
\beq
\mathcal{L}_{KN\Lambda} &=& -\frac{g_{KN\Lambda}}{M_{N}+M_{\Lambda}}\bar{\Lambda}\gamma^{\mu}\gamma_{5}\partial_{\mu}K^{-}N + h.c.\, ,\\
\mathcal{L}_{KN\Sigma} &=& -\frac{g_{KN\Sigma}}{M_{N}+M_{\Lambda}}\bar{\Sigma}\gamma^{\mu}\gamma_{5}\partial_{\mu}K^{-}N + h.c.\, ,\\
\mathcal{L}_{K\Sigma^{*} N}
&=&g_{K\Sigma^{*} N}\bar{\Sigma^{*}}_{\mu}\partial^{\mu}K^{-}N + h.c.\, ,\\
\mathcal{L}_{K^{*}N\Lambda}
&=&-g_{K^{*}N\Lambda}^{V}\bar{\Lambda}\gamma^{\mu}K^{*-}_{\mu}N+\frac{g_{K^{*}N\Lambda}^{T}}
{M_{\Lambda}+M_{N}}\bar{\Lambda}\sigma^{\mu\nu}\partial_{\nu}K^{*-}_{\mu}N + h.c.\, ,
\eeq
where notations for various symbols are standard as defined in Ref.~\cite{Nam:2006cx}. Here, we have adopted the pseudvector coupling for the kaon vertices. The approximate equivalence to the pseudoscalar type was discussed in Ref.~\cite{Nam:2003uf}.

There are seven electromagnetic interactions as given by 
\beq
\mathcal{L}_{\gamma KK} &=& -ie[(\partial_{\mu}K^{+})K^{-}-(\partial_{\mu}K^{-})K^{+}]A^{\mu}\, ,\\
\mathcal{L}_{\gamma KK^{*}}
&=&g_{\gamma KK^{*}}\epsilon_{\mu\nu\sigma\rho}(\partial^{\mu}A^{\nu})[(\partial^{\sigma}K^{+})K^{*-\rho}+(\partial^{\sigma}K^{-})K^{*+\rho}\, ,\\
\mathcal{L}_{\gamma NN} &=& -e\bar{N}(\gamma^{\mu}-\frac{\kappa_{N}}{2M_{N}}\sigma^{\mu \nu}\partial_{\nu})NA_{\mu}\, ,\\
\mathcal{L}_{\gamma \Lambda \Lambda} &=& \bar{\Lambda}\frac{e\kappa_{\Lambda}}{2M_{N}}\sigma^{\mu \nu}\partial_{\nu}\Lambda A_{\mu}\, ,\\
\mathcal{L}_{\gamma \Sigma \Lambda} &=& \bar{\Lambda}\frac{e\kappa_{\Lambda \Sigma}}{2M_{N}}\sigma^{\mu \nu}\partial_{\nu}\Sigma A_{\mu}\, ,\\
\mathcal{L}_{\gamma\Sigma^{*} \Lambda}
&=&-i\sqrt{\frac{2}{3}}(-\frac{1}{\sqrt{3}})\frac{3eg_{M}}{2M_{N}(M_{N}+M_{\Delta})}\epsilon^{\mu\nu\sigma\rho}
\bar{\Lambda}(\partial_{\mu}\Sigma^{*}_{\nu})(\partial_{\sigma}A_{\rho})\, ,\\
\mathcal{L}_{\gamma K N \Lambda} &=& ieg_{KN \Lambda} \bar{ \Lambda} \gamma^{ \mu} \gamma_{5}K^{-}NA_{\mu}\, .
\eeq
Here, various electromagnetic couplings are given as follows. The anomalous magnetic moments  are $\kappa_{p}=1.79$, $\kappa_{\Lambda}=-0.613$, 
$\kappa_{\Lambda \Sigma}=-1.61$. 
In Eq. (13) $g_{M}$ is a dimensionless coupling constant for
the $\pi N\to\Delta$ magnetic transition, $g_{M}=3.02 $~\cite{Pascalutsa:2006up}. The coupling constant of $g_{\gamma K K^{*}}$ in Eq. (8) is taken to be $0.254\ \rm{GeV}^{-1}$ in order to reproduce the 
radiative decay $K^{*\pm} \to K^{\pm}\gamma$~\cite{Yao:2006px} .
We introduce the contact $\gamma KN \Lambda$ interaction in order to preserve gauge invariance. 

As usual, we need to include the form factors for which we adopt the 
gauge invariant and covariant form factors~\cite{Ohta:1989ji,Haberzettl:1998eq}.  
\beq
F_{x}
&=&\frac{\Lambda_{c}^{4}}{\Lambda_{c}^{4}+(x-m^{2})^{2}} \    ,\; \; \;  x=s,u,t 
\label{f}.
\eeq
The cutoff parameter $\Lambda_{c}$ is commonly used for all types 
of form factors and is fixed in order to reproduce the absolute
values of the cross sections.  
Having those setups, the computation of various cross sections is 
straightforward.  

Obviously, the present choice of the Born diagrams are restricted, 
since we do not include 
nucleon resonances which are important near the threshold 
region~\cite{Lee:1999kd}.  
However, the data show that the energy dependence becomes 
rather smooth in the energy region 
$\sqrt{s} \gsim 2$ GeV, implying that the individual resonance 
effects are averaged and the background contributions become 
important.  
This is the energy region that we study in this work 
and is where the LEPS data covers~\cite{Sumihama:2005er}.  

\begin{figure}[tbp]
\centerline{
\includegraphics[width=65mm,bb=0 10 800 600,clip]{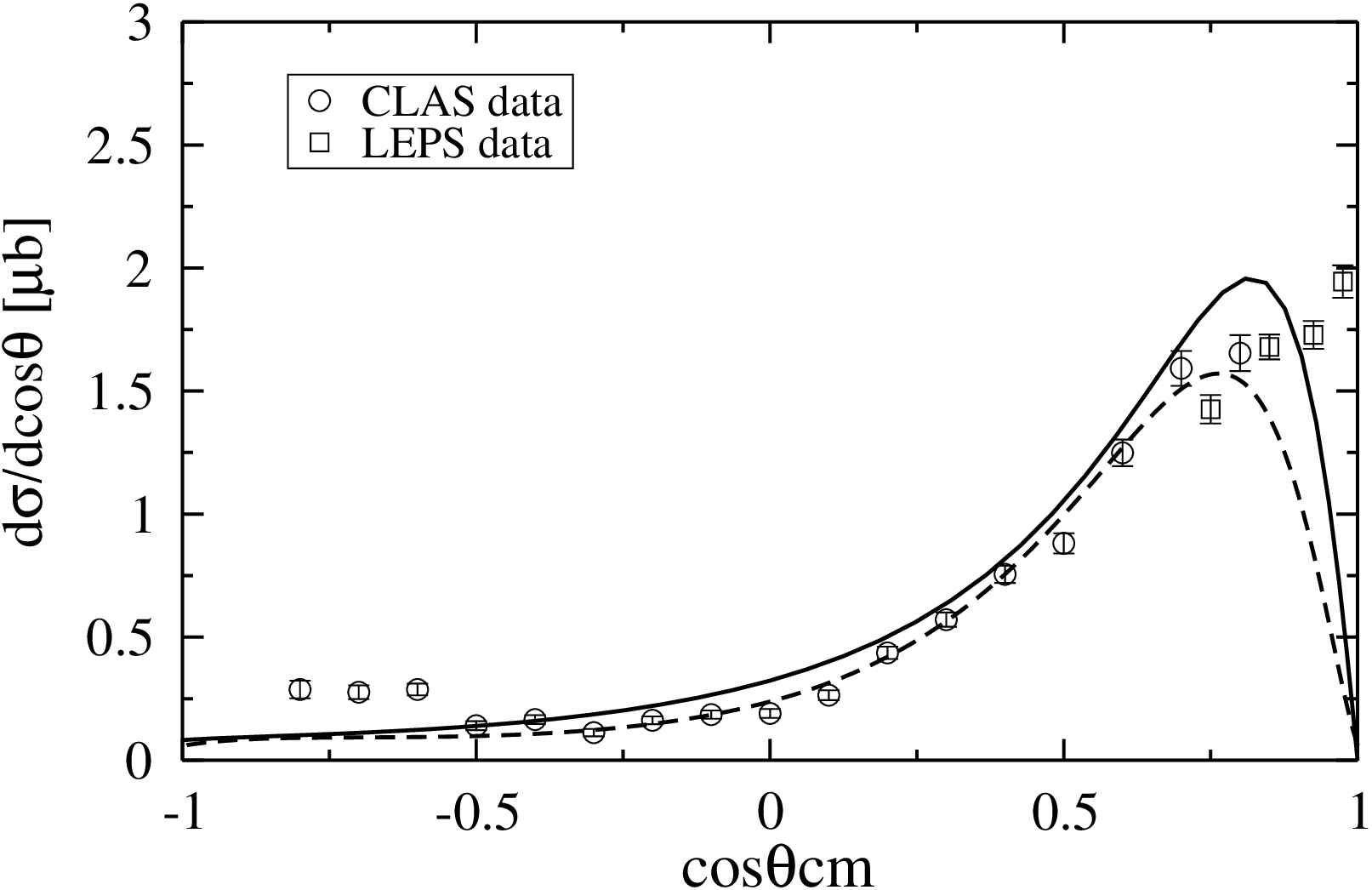}
}
\caption{\small Differential cross-sections as functions of $\cos\theta_{\rm{cm}}$at $W=2.164\ \rm{GeV}$. The solid line is the full result
including all the diagrams (a)-(e), while the dashed line shows the result of the Born diagrams (a)-(d).
The data are taken from CLAS (circle)~\cite{Bradford:2005pt} and 
from LEPS (squared)~\cite{Sumihama:2005er}. }
\label{angle}
\centerline{
\includegraphics[width=65mm,bb=0 10 800 600,clip]{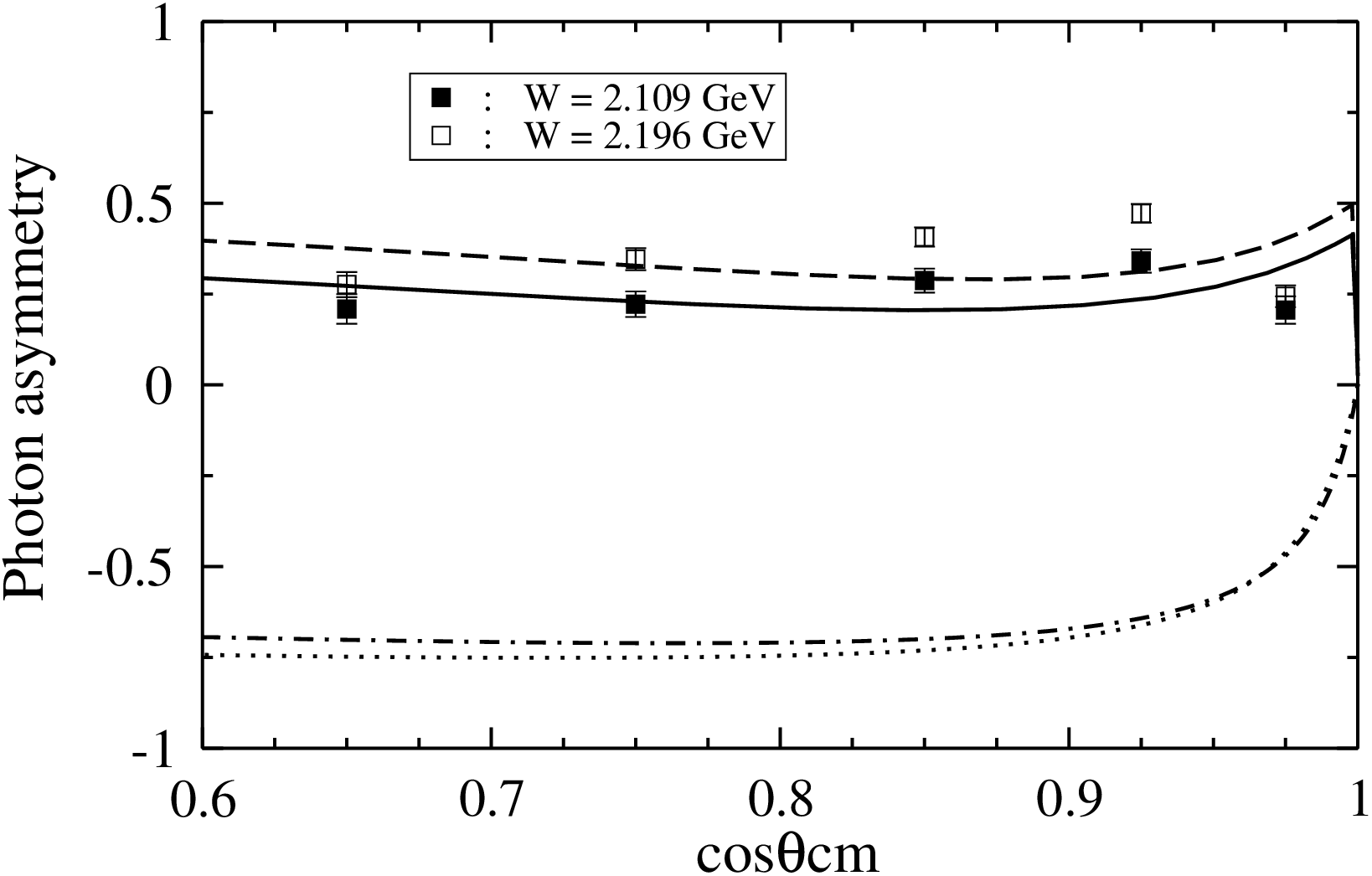}
 }
\caption{\small Photon asymmetries $A$ as functions of $\cos\theta_{\rm{cm}}$. The calculational results without the WZW
term take negative values as shown by the dotted line ($W=2.109\ \rm{GeV}$) and the dot-dashed line ($W=2.196\ \rm{GeV}$) . The full results with the WZW term are shown by the solid line ($W=2.109\ \rm{GeV}$) and dashed line ($W=2.196\ \rm{GeV}$). The data are taken from LEPS~\cite{Sumihama:2005er}.}
\label{asymmetry}
\end{figure}
%%%%%%%%%%%%  observable discussion  %%%%%%%%%%%%%%
In order to show the quality of our calculation,
we first show the differential cross section
$d\sigma/d\cos\theta$ at $W = 2.164$ GeV in Fig.~\ref{angle} as
compared with the experimental data from CLAS~\cite{Bradford:2005pt} 
and LEPS~\cite{Sumihama:2005er},
where $W=\sqrt{s}$ is the total energy in the center of mass system.
Here we first discuss the result of the Born diagrams which is
denoted by the dashed line.
In comparison with data,
the choice of the cutoff parameter is important, and
we set $\Lambda_{c} = 0.88$ GeV.
We find that the agreement is good already at the tree level.
There is some disagreement in the extremely forward
region and in the backward region.
Effects which are not included in the present study might be
important such as Reggeon
contributions~\cite{Guidal:2003qs,Corthals:2007kc,Mart:2004au}, 
coupled channel effects~\cite{Usov:2005wy,Shklyar:2005xg,Chiang:2001pw} and resonance contributions.

Now if we apply the same Born diagram calculations to
the asymmetry, as already anticipated, we fail to reproduce the positive
values as shown in Fig.~\ref{asymmetry}, where the LEPS data~\cite{Sumihama:2005er}
at $W = 2.109$ GeV (dotted line)
and at $W = 2.196$ GeV (dash-dotted line) are shown.
We have checked that the negative values are caused by the kaon exchange dominance which is of electric nature.
We would like to emphasize once again that the relatively small
$K^*$ coupling constants which are determined microscopically
are not compatible with the large magnetic interaction as required
in experiments.

Now let us consider the loop contribution induced by the WZW term.
The loop integral is proportional to
\beq
\int \frac{d^{4}q}{(2\pi)^{4}}(\Slash{p'}
-\Slash{p}-\Slash{q})\frac{1}{(p'-p-q)^{2}-m^{2}_ 
{\pi}}(\frac{\Slash{p}
+\Slash{q}-M_{B}}{(p+q)^{2}-M_{B}})\Slash{q}(\frac{1} 
{q^{2}-m^{2}_{K}})
\epsilon^{\mu\nu\sigma\rho} 
\epsilon_{\mu} q_{\nu} k_{\sigma} k'_ 
{\rho}\, ,
\eeq
where the external momenta $k, k^\prime, p, p^\prime$ are  
for the
incoming photon, outgoing kaon, incoming proton and
outgoing $\Lambda$, respectively, as
shown in Fig.~\ref{diagrams}, and $M_B$ denotes the mass of the baryon
running in the loop (either proton or $\Sigma$).
We have performed the integral by introducing the Feynman parameters
in the dimensional regularization.
After subtracting the $1/\epsilon$ term as well as constant terms
and introducing a scale
parameter $\mu$ in the form of $\ln \mu$,
we have estimated the integral numerically.
The details of computation will be reported elsewhere~\cite{ozaki}.
It is noted that due to the presence of the $\epsilon_ 
{\mu \nu \sigma \rho}$
tensor, the loop integral contributes to the same components of the amplitude as the $K^*$-exchange does.
In this sense, the loop diagram effectively renormalizes the $K^*$
coupling constants.

If we take the scale parameter $\mu$ at a hadronic scale $ 
\sim 1$ GeV,
the loop diagram brings a significant contribution to the magnetic  
interaction.
In practice, we multiply an overall form factor $F_t$ to the one loop Fig.~\ref{diagrams}(e) term
with slightly different cutoff parameter $\Lambda_{c} = 0.69$ GeV
in order to obtain a good agreement with the experimental data.
This smaller cutoff parameter is also employed for other form factors
for the Born diagrams.
It is somewhat unpleasant that the form factor is included  
here also.
We shall not, however, discuss this important issue which is
microscopically related to hadron structure, but simply follow the
empirically successful prescription.

Now we see that the effect of the loop contribution is sufficiently  
large even to flip
the sign of the asymmetry as shown in Fig.~\ref{asymmetry}.
The agreement with the experimental data is remarkable including
the increasing tendency as the photon energy is increased.
We have also verified that the differential cross section with the inclusion of
the loop diagram agrees well with the data as shown in
Fig.~\ref{angle} by the solid line.
The achievement of the good agreement for both the asymmetry
and the differential cross section is not very trivial, while varying
a single parameter $\Lambda_{c}$ in the present study.

In conclusion, we have shown that the large magnetic interaction as
observed in polarized photon experiments for the kaon photoproduction
can be qualitatively described by the inclusion of the loop diagrams
induced by the WZW term associated with QCD anomaly.
This may explain the origin of the large $K^*$ couplings which have been
phenomenologically necessary but significantly different from
what are expected in a microscopic derivation.
The present result encourages us to use the microscopic model
for various meson-baryon coupling constants which are employed in the
effective Lagrangian approach.
Having the framework based on a microscopic description will be useful
not only to the conventional reactions as discussed here
but also to extensions to more exotic phenomena.

We would like to thank Takashi Nakano for discussions.
H.~N. is supported by Research Fellowship of JSPS 
for Young Scientist and supported by Grants-in-Aid for scientific
research of JSPS  (No.~18-8661).
A.H. is supported in part by the Grant for
Scientific Research ((C) No.19540297) from the Ministry of
Education, Culture, Science and Technology, Japan.

\baselineskip 5mm

\end{document}